\documentclass[12pt]{article}
\usepackage{epsfig}
\usepackage{graphicx}
\usepackage{amssymb}      

\evensidemargin=\oddsidemargin  

\newcommand{\beq}{\begin{equation}}
\newcommand{\eeq}{\end{equation}}
\newcommand{\bq}{\begin{equation}}
\newcommand{\eq}{\end{equation}}
\newcommand{\ba}{\begin{array}}
\newcommand{\ea}{\end{array}}
\newcommand{\beqa}{\begin{eqnarray}}
\newcommand{\eeqa}{\end{eqnarray}}

\def\noi{\noindent}
\def\End{\end{document}}

\def\to{\rightarrow}

\def\[{\left[}
\def\]{\right]}
\def\({\left(}
\def\){\right)}

\def\ra{\rightarrow}

\def\U1EM{U(1)_{\rm em}}

\def\[{\left[}
\def\]{\right]}

\begin{document}
\hfill {WM-03-107}

\hfill {\today}

 \baselineskip 24pt
\setcounter{footnote}{0}   
\input epsf

\begin{center}
 {\Large \bf $\mu + N \ra \tau + N$ at a Muon or Neutrino Factory }
\end{center}
\vskip 1.0cm
\begin{center}
    { {\sc Marc Sher and Ismail Turan \footnote{Email: sher@physics.wm.edu}}\\ Nuclear and Particle Theory Group\\Physics Department\\ College of
William and Mary, Williamsburg, VA 23187, USA}

\end{center}
\date{(~\today  \,~and~ hep-ph/0207136~)}  

\vspace{24pt}

\begin{abstract}
\noindent  The experimental discovery of large $\nu_\mu-\nu_\tau$ mixing
        indicates that analogous mixing in the charged lepton sector could be substantial.  We
        consider the possibility that if a high intensity muon beam, perhaps at the early stages of a
        muon or neutrino factory, strikes a nuclear target, then conversion of some of the muons into
        tau leptons could occur (similar to the conversion of muons to electrons at MECO).  Using
        current experimental limits on rare tau decays to bound the size of the relevant operators, we
        find that a $50$ GeV muon beam, with $10^{20}$ muons on target per year, could yield as many
        as $10^7$ $\mu+N\to\tau+N$ events per year.  Backgrounds could be substantial, and we comment
        on the possibility of detection of this process.

\end{abstract}

\newpage
\noi In the past decade, the biggest surprise in our understanding of flavor physics has been the
        discovery of large mixing \cite{neutrino} in the neutrino sector.  This large mixing may come from
        diagonalization of the neutrino mass matrix, the charged lepton mass matrix, or both.   It is
        quite possible that searches for charged lepton flavor violation will be critical in
        determining the physics of flavor violation.

        The most promising such search is the MECO experiment \cite{meco}, in which a high intensity, low energy
        muon beam strikes a nuclear target.  The muons are captured, and decay essentially at rest. 
        The conversion $\mu + N \to e + N$ will then yield a distinctive $105$ MeV electron.   The
        experiment promises to achieve the extraordinary sensitivity of a part in $10^{17}$, and in
        many models beyond the standard model, a positive signal would be expected.

        Observations of atmospheric neutrinos indicate that mixing between the muon and tau neutrinos
        is maximal.   This gives strong motivation for considering transitions between the muon and
        the tau.  Of course, the analogous process to MECO, $\tau + N \to \mu + N$, is impractical due
        to the short lifetime of the $\tau$.   However, the inverse process, $\mu + N \to \tau + N$
        might be possible.  Unlike MECO, this can't occur for muons at rest, but in a higher energy
        muon beam, one can look for such events.  Such high energy and high intensity muon beams are
        expected \cite{nufact} at neutrino factories (or early stages of muon factories), in which intensities of
        $10^{20}$ muons per year and beam energies up to $50$ GeV have been proposed.    In this note,
        we examine whether the $\mu + N \to \tau + N$ process is feasible at such a neutrino factory.

        The existence of the process $\mu + N\to \tau + N$ immediately implies that there will be muon
        and tau number violating rare $\tau$ decays, such as $\tau\to\mu\pi, \tau\to\mu\pi\pi,
        \tau\to\mu\rho, $ etc.    The non-observation (as yet) of these decays implies an upper bound
        on $\mu + N \to \tau + N$.   We first examine the upper bound on the size of the various
        operators.

        The relevant operators are of the form $(\bar\mu\,\Gamma\,\tau)(\bar{q}^\alpha\,\Gamma\, q^\beta)$,
        where $\Gamma$ contains various combinations of Dirac gamma matrices.   A detailed analysis of
        all 48 possible operators, where the $q$'s are any combination of the six quarks and the
        $\Gamma$ consists of $(1,\gamma_5,\gamma_\mu,\gamma_\mu\gamma_5)$  was carried out recently by
        Black, et al. \cite{black}.   They determined the experimental lower bound on $\Lambda$ for each process,
        where $\Lambda$ is defined by the considering the relevant operator to be 
        \begin{eqnarray}
        {4\pi\over\Lambda^2}(\bar\mu\,\Gamma\,\tau)(\bar{q}^\alpha\,\Gamma \,q^\beta).
        \end{eqnarray}

        For simplicity, we will consider valence quarks only, and will assume that the operators are
        isospin invariant, so that the operators involving $u$-quarks and $d$-quarks are the same
        magnitude.  Relaxing this assumption will only strengthen our results.  Black, et al. find
        that the lower bound on $\Lambda$ for $\Gamma=(1,\gamma_5,\gamma_\mu,\gamma_\mu\gamma_5)$ is
        $(2.6, 12, 12, 11)$ TeV, which come from $\tau\to\mu\pi^+\pi^-,
        \tau\to\mu\pi^o, \tau\to\mu\rho$ and $\tau\to\mu\pi^o$, respectively.   Since the bound on the
        scalar operator is the weakest, we will assume that the operator is scalar, and is thus
        \begin{equation}     
        {4\pi\over\Lambda^2}(\bar\mu\,\tau)(\bar{q}\, q),
        \end{equation}    
 where $q$ is $u$ or $d$ and $\Lambda$ is
        greater than $2.6$ TeV.    No experiment can currently exclude such a possibility.  When our
        results are presented, we will briefly comment on the effects of choosing one of the other
        three operators.

        With this operator, we can calculate the cross section for $\mu + q \to \tau + q$, and we find
        that 
        \begin{equation} 
        \sigma(\mu + q \to \tau + q) = \left({\pi s\over 3\Lambda^4}\right) \left( 1-{m^2_\tau\over s}\right)^2\left(1+{m^2_\tau\over 
         2s}\right).
        \end{equation}
  Folding in the parton distribution functions, we plot the cross
        section for $\mu + N \to \tau + N$, where $N$ is a nucleon, in Figure 1, assuming 
        that the lower bound on $\Lambda$ is saturated.   For the expected 
        beam energy of 50 GeV, the cross section is $0.55\  {\rm fb}$.

\begin{figure}[htb]
    \centerline{ \epsfxsize 3.0in {\epsfbox{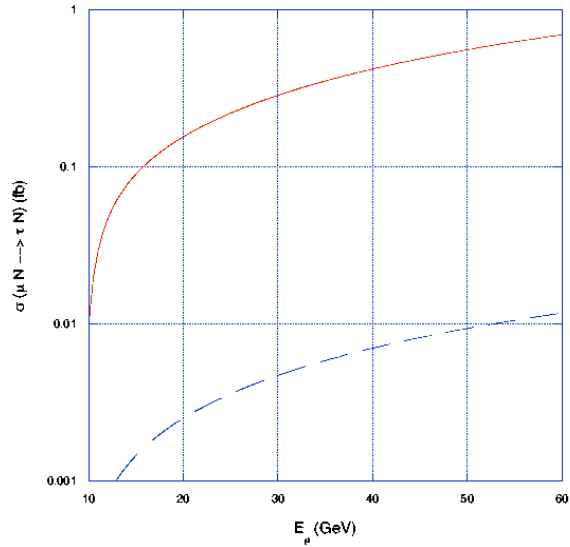}}}
    \caption{The cross section for the scattering $\mu\,N\rightarrow \tau\,N$ 
      in units of fb as a function of muon energy (GeV) in Lab frame. The solid (dashed)
      line represents the cross section assuming a scalar (vector) interaction.}\label{fig}
    \end{figure}

With this cross section, we can determine the mean free path.  If $\rho$ is the density of the
        target (in g/cm${}^3$), the mean free path is 
        \begin{equation}  
 \lambda = {1\over \rho}\left({1\ 
        {\rm fb}\over\sigma}\right) (1.6\times 10^{13})\ {\rm meters}.
        \end{equation}   
 For a 50 GeV muon beam, there is little ionization
        loss over a meter of target, and thus there is a probability of approximately $3\times 10^{-14}\rho$ 
        of interacting
        in a meter of target.  With $10^{20}$ muons on target is a year, 
this gives $3\times 10^6\rho$ events
        per year per meter of target.
        
        We have assumed that the interaction is scalar.  If it is vector, there is a factor of $8$ increase in the square of 
        the matrix element (in the massless limit),  however
        the lower bound on $\Lambda$ is $12$ TeV instead of $2.6$ TeV, leading to a lower event rate.
        This is also plotted in Figure \ref{fig}.   Nonetheless, even here there could be well over $100,000$ events 
        per year.   Using pseudoscalar or axial vector operators will give similar results.  But for the scalar
        case, and a fairly dense target, the event rate could exceed $10^7$ events per year.

        Although this seems to be a huge event rate, the backgrounds could be severe.  Note, however
        that the cross section 
        for tau pair production through Bethe -- Heitler production off iron nuclei \cite{ganapathi}  is much smaller 
        than a femtobarn, and the $p_T$ distributions are much softer, so tau pair
        production will not be a problematic background.    The major difficulty is identifying a clear
        signature.  A typical $\tau$
        energy, for a $50$ GeV incident beam energy, will be tens of GeV, and thus its decay distance 
        would be a couple of millimeters.   One can imagine alternating target and scintillator, but many $\tau$'s will be
        missed.  The only places in which $\tau$'s have been detected are the clean environment of electron-positron colliders, the Tevatron, where the signature is large missing transverse energy, and DONUT \cite{donut}.  The latter used 
        lead and emulsions
        as the target and detection media, and it isn't clear whether the enormous intensity of the incident
        muon beam would blacken the emulsion (this would depend on the beam size and whether the emulsion is cycled in and out).
           This possibility should be investigated.  
        What are the specific decay modes that might be observable?
        The leptonic decays will clearly be swamped by backgrounds.   The $\pi\nu$ decay mode
        will lead to a monochromatic pion, but unless a $\tau$ track can be observed, the backgrounds for 
        single pions in the intense muon beam will also be very large.  One could look at rarer decays,
        such as the three charged pion (or even five charged pion) decays, coming at the end of a very
        short track.  Clearly, detection of this process will not be easy, but the event rate is high enough  
        that a clever scheme might be able to pick out a signal.

           Are there specific models which predict such a large rate for $\mu + N 
\rightarrow \tau + N$?  The Standard Model, with massive neutrinos, 
will have mixing between the $\mu$ and the $\tau$, but this mixing is 
of the order of $m_\nu^2/m_W^2$, and is thus negligible.  However, 
there are a wide variety of extensions of the Standard Model, including 
models with very heavy neutrinos, horizontal symmetries, left-right 
symmetry, supersymmetry, extended gauge and Higgs models, etc., and 
many of these do predict such mixing to occur at a substantially higher 
rate.  The effects of $\mu-\tau$ mixing can be parametrized by 
operators of the form of Eq. (1).  As noted earlier, the biggest rates for
$\mu + N \rightarrow \tau + N$ occur if the operator is scalar, as in 
Eq. (2) (since the experimental limits on the operator are weaker), and 
thus models with flavor-changing scalar exchanges are most promising.  
For example, in R-parity violating supersymmetry \cite{gad}, the 
superpotential can be written in the form $\lambda_{ijk} L_i L_j E_k + 
\lambda^\prime_{ijk} L_i Q_j D_k$.  If the underlying theory giving 
rise to this superpotential gives a hierarchical structure for 
$\lambda$, so that $\lambda_{i23}$ is large, and a non-hierarchical 
structure for $\lambda^\prime$, then the operator of Eq. (2) can be 
generated via scalar neutrino exchange.  If the couplings are of order 
unity, and the scalar neutrino mass is of the order of a TeV, then the 
operator will be as large as allowed by bounds on 
$\tau\rightarrow\mu\pi\pi$ and the rate for $\mu + N \rightarrow \tau + 
N$ will be as large as discussed in the previous paragraph.  
Alternatively, supersymmetric models at large $\tan\beta$ can have very 
large flavor-changing Higgs couplings \cite{babu}, and that can also 
lead to similarly large muon to tau conversion.  Thus, we see that 
plausible extensions of the Standard Model exist in which 
$\tau\rightarrow\mu\pi^{+}\pi^{-}$ is near its current limit,  

           The early stages of a neutrino or muon factory will involve a high intensity muon beam with
        energies up to $50$ GeV.  In this Brief Report, we have proposed that such a facility may be able to 
        substantially improve bounds on $\mu-\tau$ mixing, or discover such mixing, by looking for muon 
        conversion in nuclei to tau leptons.  The event rate could be high, although backgrounds will 
        be challenging.  In view of the large
        mixing in the neutrino sector, this may be a promising place to search for new flavor physics.
        
	After this work was completed, we became aware of a very interesting paper by Gninenko, Kirsanov, Krasnikov 
and Matveev \cite{gnin}.  They also considered the process $\mu + N \to \tau + N$ at a neutrino factory.  
Instead of considering the vertex involving valence quarks, as we did, they considered the four-fermi interaction 
$(\bar{\mu}\tau)(\bar{u}c)$, involving production of a charmed quark.  This has a substantial advantage over our vertex,
 which is flavor diagonal, because there are no experimental constraints on the size of this interaction 
(since $\tau$'s can't decay into a charmed meson plus a muon).  As a result, they had a much higher event rate, 
and could consider the muonic decay of the $\tau$.  They performed a simulation of the signature at the NOMAD detector.
	What is new in our work?  We considered the flavor diagonal vertex, which is more tightly constrained by 
experimental bounds.  Our belief is that it is very unlikely for the four-fermi interaction to be purely off-diagonal in the mass 
eigenstate basis, and thus the existence of the vertex considered by Gninenko et al. will generally imply the existence 
of the vertex that we have considered.  In that sense, this work is complementary to theirs.  Clearly, there is sufficient 
interest in the possibility of mu-tau conversion in nuclei that all experimental possibilities should be considered.

        We are very grateful to Jon Urheim for numerous discussions, and we also appreciate discussions
        with Andy Sessler, David Armstrong, Stan Brodsky, Jack Kossler and Jeff Nelson.  The work of MS was supported by
        the National Science Foundation under grant PHY-0243400, and the work of IT was supported by 
      the Scientific and Technical Research Council of Turkey (T\"{U}B\.ITAK) in the framework of
      NATO -- B1 program.


\end{document}